\documentclass[useAMS,usenatbib]{mn2e}

\usepackage{epsfig}
\usepackage{graphicx}
\usepackage{rotating}
\usepackage{hyperref}

\title[Classifying BALQSOs: Metrics, Issues and a New catalogue]
{Classifying Broad Absorption Line Quasars: Metrics, Issues and a New Catalogue Constructed from SDSS DR5}
\author[S. Scaringi \textit{et al.}]
{S. Scaringi$^{1}$\thanks{E-mail: simo@astro.soton.ac.uk}, C.E. Cottis$^{2}$, C.Knigge$^{1}$, M.R. Goad$^{2}$\\ $^{1}$Department of Physics and Astronomy, University of Southampton, Highfield, SO17 1BJ, UK \\ $^{2}$Department of Physics and Astronomy, University of Leicester, University Road, LE1 7RH, UK}

\begin{document} 

\date{}

\pagerange{\pageref{firstpage}--\pageref{lastpage}} \pubyear{2009}
\maketitle

\label{firstpage}

\begin{abstract}
We apply a recently developed method for classifying broad absorption line quasars (BALQSOs) to the latest QSO catalogue constructed from Data Release 5 of the Sloan Digital Sky Survey. Our new hybrid classification scheme combines the power of simple metrics, supervised neural networks and visual inspection. In our view the resulting BALQSO catalogue is both more complete and more robust than all previous BALQSO catalogues, containing 3552 sources selected from a parent sample of 28,421 QSOs in the redshift range $1.7<z<4.2$. This equates to a raw BALQSO fraction of $12.5\%$.
    
In the process of constructing a robust catalogue, we shed light on the main problems encountered when dealing with BALQSO classification, many of
which arise due to the lack of a proper physical definition of what constitutes a BAL. This introduces some subjectivity in what is meant by the term BALQSO, and because of this, we also provide all of the meta-data used in constructing our catalogue, for every object in the parent QSO sample. This makes it easy to quickly isolate and explore sub-samples constructed with different metrics and techniques. By constructing composite QSO spectra from sub-samples classified according to the meta-data, we show that no single existing metric produces clean and robust BALQSO classifications. Rather, we demonstrate that a variety of complementary metrics are required at the moment to accomplish this task. Along the way, we confirm the finding that BALQSOs are redder than non-BALQSOs and that the raw BALQSO fraction displays an apparent trend with signal-to-noise, steadily increasing from $9\%$ in low signal-to-noise data, up to $15\%$.
\end{abstract}


\section{Introduction}

Broad absorption line quasars (BALQSOs) are a sub-class of active galactic nuclei (AGN) exhibiting strong, broad and blue-shifted spectroscopic absorption features (\citealt{foltz90,weymann91,reichard03b,hewett03}). These features are thought to be formed in fast ($0.1c$-$0.2c$) and powerful outflows from the
accretion disk around the supermassive black hole at the heart of the AGN (\citealt{korista92}). The vast majority of BALQSOs are radio-quiet (\citealt{stocke92}, but see \citealt{brotherton06} for some counter examples), and there are subtle differences between their continuum and emission line properties and those of “normal” (non-BAL) QSOs (\citealt{reichard03b}). However, despite these differences, BALQSOs and non-BALQSOs appear to be drawn from the same parent population (\citealt{reichard03b}). Most BALQSOs belong to the subclass of the so-called HiBALs, only displaying absorption in certain high-ionisation lines (e.g. N~{\sc v}~1240~\AA, C~{\sc iv}~1549~\AA, S~{\sc iv}~1397~\AA). However, some, known as LoBALs, also show absorption in some low-ionisation lines (most notably Mg~{\sc ii}~2800~\AA). 

The most straightforward explanation for the differences between QSOs and BALQSOs is a simple orientation effect. Thus \textit{all} QSOs may undergo significant mass loss through winds (\citealt{ganguly}), but BALs are only observed if the central continuum and/or emission line source is viewed directly through the outflowing material. Viewed in this context, BALQSOs may be the only available tracers of a key physical process common to all AGN. Also, the powerful outflows we observe in BALQSOs are an important example of AGN feedback in action (\citealt{tremonti}). Such feedback is a key ingredient required in theoretical attempts to understand galaxy ``downsizing'' and may also be responsible for regulating the growth of supermassive black holes. Moreover, the fraction of QSOs displaying BAL features ($f_{BALQSO}$) may provide a direct estimate of the opening angle of these outflows. 

Historically, BALQSO samples have been selected on the basis of the so-called balnicity index (BI; \citealt{weymann91}) or similar metrics. These samples consistently yielded BALQSO fraction estimates in the range $f_{BALQSO} \approx 0.10$ - $0.15$ (\citealt{weymann91,tolea,hewett03,reichard03a}). In a previous paper (\citealt{knigge08}; Paper~I), we showed that both the BI and a more recently defined metric, the absorption index (AI; \citealt{trump06}), are biased when selecting BALQSOs, the former being incomplete at the low-velocity end of the BALQSO distribution, and the latter suffering from significant contamination by objects with low-velocity absorption systems which may be unrelated to the higher velocity outflows. 

Here, we use a combination of the classic BI metric, a simple neural network and visual inspection (the hybrid-LVQ approach we developed in Paper~I) to produce a BALQSO sample that is both more complete than purely BI-based ones and, importantly, significantly more robust than AI-based ones. We have applied our hybrid-LVQ algorithm to the QSO sample associated with Data Release 5 (DR5) of the Sloan Digital Sky Survey (SDSS;\citealt{BIG_dr5,dr5}) using the BIs calculated from \cite{gibson08}. The resulting catalogue contains 3552 BALQSOs selected from a parent sample of 28,421 QSOs on the basis of absorption close to the C~{\sc iv} high-ionisation emission line. This catalogue may be obtained from \texttt{http://www.astro.soton.ac.uk/$\sim$simo}. A preliminary version of the catalogue has already been presented in \cite{scaringi08}. In addition, we also provide (at the same address) a catalogue of the meta-data, i.e. the data pertaining to the parent QSO sample and subsequently used in the compilation of our BALQSO catalogue, so that members of the scientific community wishing to compile their own BAL/non-BAL subsamples may readily do so.

\section{Data and Methods}

\subsection{The QSO parent population}
The SDSS DR5 QSO catalogue contains over 77,000 objects in total (\citealt{dr5}). However, for the purpose of constructing a uniform BALQSO catalogue, we only consider objects whose spectra fully cover the C~{\sc iv}~1550~\AA\ resonance line and its associated absorption region (up to $29000$ ${\rm km~s^{-1}}$ blueward of the C~{\sc iv} line centre), which displays a particularly deep and well-defined absorption trough in the spectra of most BALQSOs. Given the wavelength range covered by the SDSS spectra, this implies an effective redshift window of $1.7 < z < 4.2$ for our QSO parent sample. This redshift window yields spectra for a QSO parent sample of 28,421 objects. This will be the parent sample used in this study to compile our BALQSO catalogue.

\subsection{Metrics and Preconditioning} 
Our BALQSO classification method works on continuum normalised spectra covering the wavelength range 1401~\AA\ - 1700~\AA\ with 1~\AA\ dispersion. It also uses the associated BIs for training the neural network and to flag borderline cases requiring visual inspection. The BI metric is defined as
\begin{equation}
BI=-\int_{25000}^{3000}\left[1-\frac{f(v)}{0.9}\right]Cdv.\label{eq:1}
\end{equation}
Here, the limits of the integral are in units of $km~s^{-1}$ , $f(v)$ is the normalised flux as a function of velocity displacement from line centre. The constant $C = 0$ everywhere, unless the normalised flux has satisfied $f(v) < 0.9$ continuously for at least $2000$ ${\rm km~s^{-1}}$ , at which point it is switched to $C = 1$ until $f(v) > 0.9$ again. Based on this definition, objects are classified as BALQSOs if their $BI>0$ ${\rm km~s^{-1}}$ . The BI by definition excludes strong, low-velocity absorption systems; for example, any deep absorption of width $3000~{\rm km~s^{-1}}$ which starts less than $2000$  ${\rm km~s^{-1}}$ blueward of the rest wavelength of the C~{\sc iv} emission line will be assigned $BI=0$ ${\rm km~s^{-1}}$. Thus BALQSO catalogues constructed using the BI metric are likely to be significantly incomplete at the low velocity end of the distribution.

For this reason \cite{hall02} introduced the so-called AI (Absorption Index), in an attempt to recover those low-velocity absorption systems objects that were missed by the BI. The AI is defined as
\begin{equation}
AI=\int_{0}^{29000}\left[1-f(v)\right]Cdv,\label{eq:2}
\end{equation}
here now $C = 1$ in all regions where $f(v) > 0.9$ continuously for at least $1000$ ${\rm km~s^{-1}}$ and $C = 0$ otherwise. The two key differences that allow objects with $BI=0$ ${\rm km~s^{-1}}$ to achieve $AI> 0$ ${\rm km~s^{-1}}$ are (i) that the AI includes regions within $3000~{\rm km~s^{-1}}$ of line centre (and also regions beyond $25,000$ ${\rm km~s^{-1}}$ ) and (ii) that the AI includes objects with much narrower absorption troughs than the BI. The remaining differences are associated with the presence [absence] of the factor $0.9$ in Equations \ref{eq:1} and \ref{eq:2}. The less stringent constraints imposed by the AI allow one to recover the majority of the low velocity absorption systems missed by the BI, more than doubling the number of objects classed as BALQSOs. However, as shown in Paper I, the log-AI distribution is bi-modal, with low-velocity outflows preferentially occupying one mode and high-velocity outflows occupying the other. While it is beyond doubt that at least some of the BALQSOs classified solely by the AI are bona-fide BALQSOs in the traditional sense, particularly in the region where the two modes overlap, it remains uncertain whether the two modes are physically connected. Thus we cannot exclude the possibility that the AI includes substantial numbers of objects whose low-velocity absorption systems are unrelated to the high-velocity flows traditionally associated with the BALQSO phenomenon. Specific examples of hard-to-classify BALQSO spectra selected using either the AI or BI may be found in Paper~I.

\begin{figure*}
\centering
\includegraphics[height=0.8\textwidth, angle=90]{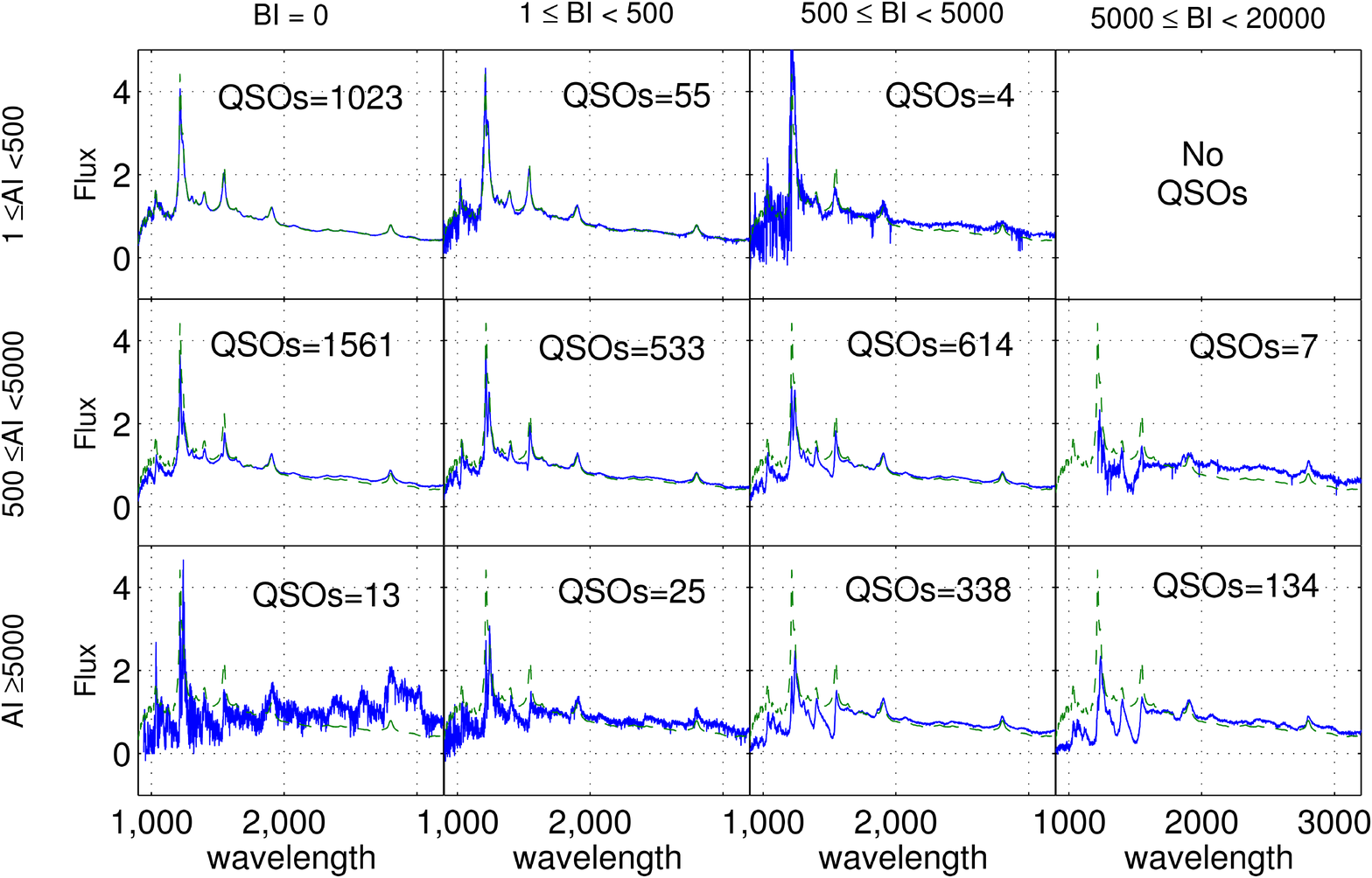}	
\caption{Composites in various AI-BI ranges (blue line), and composites created from $AI=0$ ${\rm km~s^{-1}}$ and $BI=0$ ${\rm km~s^{-1}}$ objects (dashed green line). The AI and BI bins on the side panels are in ${\rm km~s^{-1}}$. Reddening has not been taken into account.}
\label{fig:1}
\end{figure*}

The classification problems described above are illustrated in Fig. \ref{fig:1}. The figure displays QSO geometric mean composites created using the DR3 subset from our DR5 parent population normalised at 1750~\AA\ \footnote{We have used the DR3 subset so that we can use the AIs provided by \cite{trump06}}. More specifically, it shows the average properties of QSO spectra on a grid in AI/BI space, allowing a close examination of the absorption through dependence on the AI and the BI. For reference we have also included in each panel the same non-BALQSO composite created from QSOs with $AI = 0$ ${\rm km~s^{-1}}$ (dashed green curve).

It is generally clear from Fig. \ref{fig:1} that both the AI and the BI tend to select redder QSOs, and that not only do the troughs get wider with increasing AI/BI, but also deeper. Moreover, Fig. \ref{fig:1} shows how QSO samples selected from the low-velocity region of the AI do not display the ``traditional'' BALQSO properties. This is best shown in the second composite from the left panel (top) with $BI=0$ ${\rm km~s^{-1}}$ and $1$ ${\rm km~s^{-1}} AI < 500$ ${\rm km~s^{-1}}$ , which shows little, if any, sign of absorption when compared to the non-BAL composite. The next panel down  displays a composite created from 1561 QSOs (the largest sub-sample) which have $500$  ${\rm km~s^{-1}}<$ $AI < 5000$ ${\rm km~s^{-1}}$ and $BI=0$ ${\rm km~s^{-1}}$. Relative to the non-BAL composite, there is some evidence of absorption close to the C~{\sc iv} emission line. However, we caution that, since, the BAL composite is significantly redder than the non-BAL composite, identifying broad absorption lines in this spectrum is difficult, without first dereddening the spectrum. The remaining panels show spectra with increasingly prominent absorption in the vicinity of C~{\sc iv}, with the absorption strength (depth) and reddening increasing both with increasing AI (moving from top to bottom), and with increasing BI (left to right). 

We conclude from examining Fig. \ref{fig:1} that using the $AI>0$ ${\rm km~s^{-1}}$ to select BALQSOs is unreliable, since QSOs with $0$ ${\rm km~s^{-1}}$ $< AI < 5000$ ${\rm km~s^{-1}}$ and $BI=0$ ${\rm km~s^{-1}}$ have spectra that are not different from $AI=0$ ${\rm km~s^{-1}}$ non-BALQSOs. Moreover, most BALQSOs fall in the low-velocity region of both the AI and the BI continuum, which is also the region which turns out to be the hardest to classify. However, it is interesting to note that QSOs with $AI>5000$ ${\rm km~s^{-1}}$ and $BI<500$ ${\rm km~s^{-1}}$ do look like BALQSOs. We have decided to omit the AI metric from our hybrid classification method since about $\approx 50\%$ of the objects selected using this metric may not be genuine BALQSOs (see Paper~I). Instead, our hybrid-LVQ method uses the BI, a simple neural network and visual inspection to select BALQSOs.

\subsection{Hybrid-LVQ selection of BALQSOs} 

The method we use to classify BALQSOs has already been described in detail in Paper~I, so we only provide an overview of the key points here. Briefly, our method is a hybrid of BI-based, neural network and visual classifications, and is designed to produce a more complete BALQSO sample than a pure BI selection,  but without significantly increasing the number of false positives. Starting with a BI-based classification (as calculated by \citealt{gibson08}), we use a simple neural network-based machine learning algorithm called ``learning vector quantization'' (LVQ, \citealt{kohonen01}) to identify objects that might have been miss-classified by the BI. All such objects are then inspected and classified visually. 

We caution that both the measured BI (and the AI for the reference) are very sensitive to our ability to perform an accurate fit to the underlying continuum. Overestimating the underlying continuum strength can yield a large positive AI and BI in the absence of any absorption. Conversely, if the continuum is underestimated, weak broad absorption features may go unrecognised. This is an issue which can also affect our hybrid-LVQ classification method. For this reason we have decided to use the BI’s calculated from \cite{gibson08}, since their continuum fitting algorithm is likely to be superior to the one we use for normalising spectra for input into LVQ. This is mainly because they employ multiple composites in order to fit the underlying continuum (\citealt{trump06}). 

For input into LVQ, we normalise all QSO spectra using the method described in \cite{knigge08} and \cite{north06}, in which each spectrum is fitted with a modified DR3 QSO composite (constructed from objects with $AI=0$ ${\rm km~s^{-1}}$ as calculated from \citealt{trump06}) allowing for object-to-object differences in reddening and spectral index. We then bin each spectrum onto a uniform grid in wavelength, and use the binned spectrum between 1401\AA\ - 1700\AA\ for our classification purposes. 

The way we train our LVQ network to recognize BALQSOs has been described in detail in Paper~I. In brief, we employ a training set composed of 400 $BI>0$ ${\rm km~s^{-1}}$ and 400 $BI=0$ ${\rm km~s^{-1}}$ QSOs and train our LVQ-network to recognise $BI>0$ ${\rm km~s^{-1}}$ objects at first. We then visually inspect our neuron map for BALQSO mis-classifications (locating $BI>0$ ${\rm km~s^{-1}}$ QSOs in $BI=0$ ${\rm km~s^{-1}}$ nodes and vice versa) and re-tag those objects. We then retrain our LVQ-network using the new BALQSO vs. non-BALQSO tags to create a final neuron map. Note that redshift uncertainties are explicitly taken into account by our network and all the spectra have been de-reddened to match the non-BALQSO composite. Below, we will sometimes refer to the full hybrid method as “LVQ-based”, but it is always worth keeping in mind that LVQ is only one part of a process also involving the BI and visual inspection.

\subsection{The final BALQSO catalogue} 

\begin{figure}
\centering
\includegraphics[width=0.5\textwidth, height=12cm]{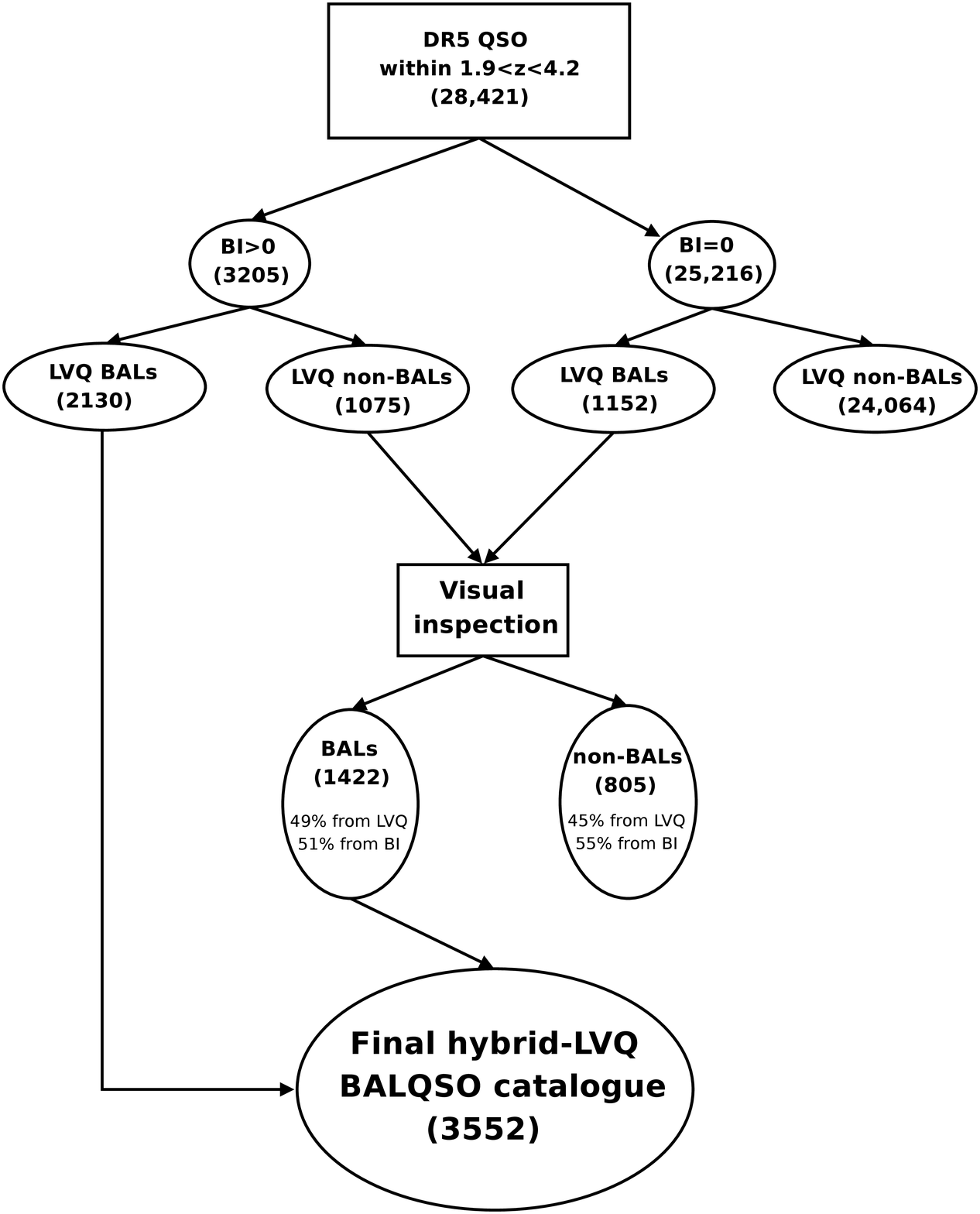}	
\caption{Flow diagram illustrating the steps involved in our hybrid-LVQ classiﬁcation method.}
\label{fig:2}
\end{figure}

Our LVQ based DR5 BALQSO catalogue contains 3552 objects ($\approx 12.5\%$ of the parent QSO sample). Fig. \ref{fig:2} shows a flow diagram detailing the individual steps involved in creating this catalogue, along with the numbers of QSOs associated with each step. Overall, we find that 3205 QSOs ($11.3\%$ of the parent sample) are classified as BALQSOs by the BI metric (i.e. $BI > 0$ ${\rm km~s^{-1}}$), and 3282 QSOs ($11.5\%$) are classified as BALQSOs by the LVQ network alone (without visual inspection). The subset of objects classified as BALQSOs by both methods comprises 2130 QSOs ($7.5\%$), and only these are added to the final catalogue without undergoing visual inspection. All of the QSOs for which the BI and LVQ classifications disagree are inspected and classified visually. This step contributes a further 1422 objects ($5.0\%$) to the catalogue. 

That BALQSO classification can be difficult is highlighted when one considers the percentage of false identifications produced by each of the two automated methods (i.e. the BI and LVQ) in isolation. The 3552 BALQSOs in our final catalogue include 2840 of the 3205 objects classified as BALQSOs by the BI metric calculated by \cite{gibson08}. Thus the BI alone would have missed $20.0\%$ (712/3552) of the objects in our final catalogue and produced false positives at a rate of $11.4\%$ (365/3205). Similarly, LVQ alone would have missed $20.0\%$ (710/3552) of our BALQSOs and produced false positives at a rate of $13.4\%$ (440/3282). Both methods individually yield very comparable false identification rates, which highlights the large uncertainties associated with previous BALQSO classifications. Clearly, designing a fully automated, reliable and reasonably complete classification scheme for BALQSOs is a difficult task. 

\begin{figure*}
\centering
\includegraphics[width=\textwidth, height=0.6\textheight]{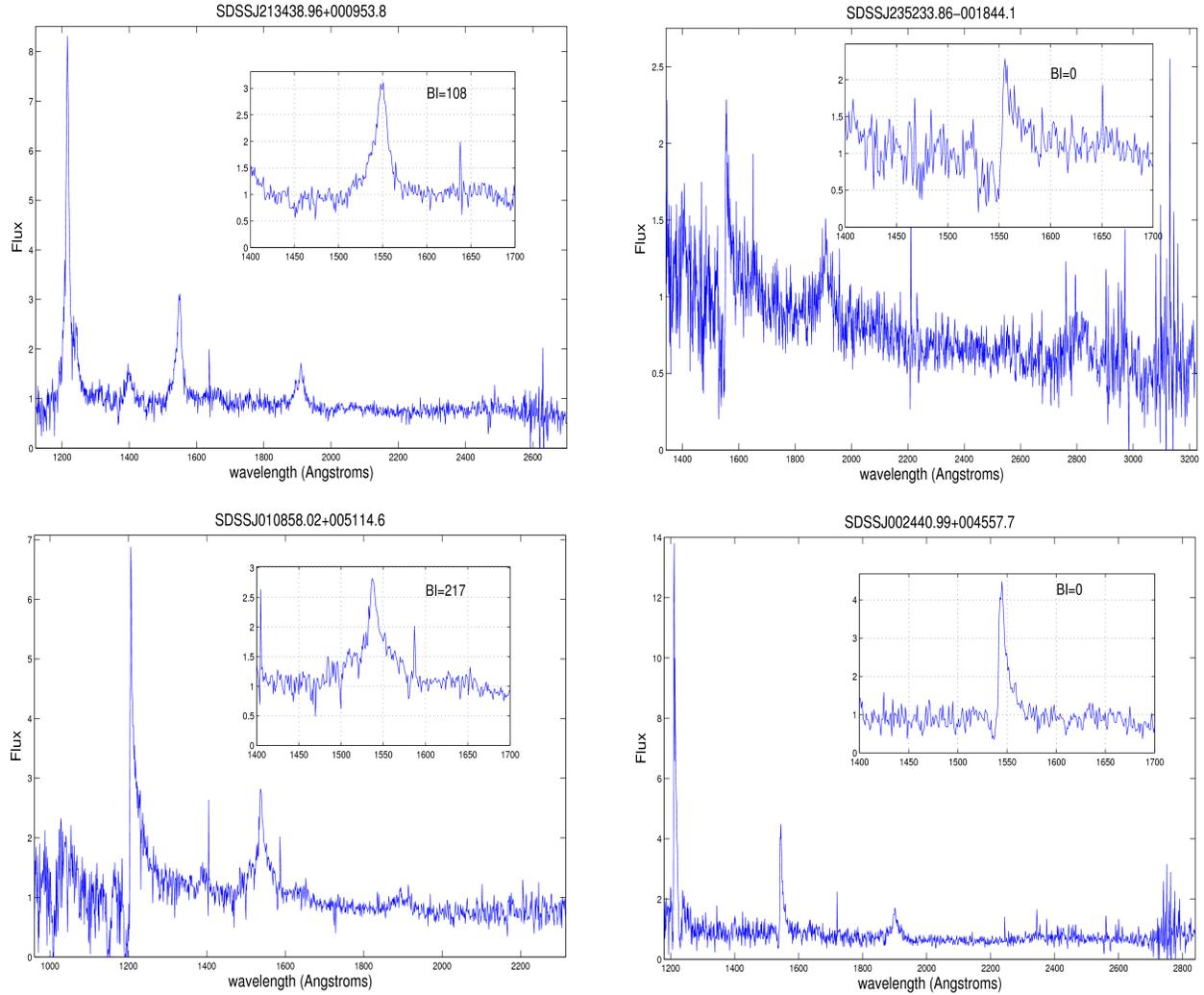}	
\caption{Four QSO spectra with different classification tags from the \protect\cite{gibson08} catalogue and ours. The two spectra on the left have positive BI's but are not included in our BALQSO catalogue, whilst the two spectra on the right have $BI=0$ ${\rm km~s^{-1}}$ and are included in our catalogue.}
\label{fig:3}
\end{figure*}

In order to explore this issue further, we present Fig \ref{fig:3}, which shows four QSO spectra that highlight some of the subjectivity and difficulty associated with classifying BALQSOs. The two spectra on the left side both have positive BI’s and, as a consequence, are included in the \cite{gibson08} BALQSO catalogue, but not in ours. These were objects which were tagged as non-BALQSOs by our LVQ network and were visually inspected for final classification, since they both had positive BI’s. The spectrum in the bottom left (SDSS J010858.02$+$005114.6) provides a particularly useful insight. Here, the C~{\sc iv} line shows no sign of absorption, and thus this object was not classified as a BALQSO by us. However, there is some evidence that the Lyman alpha line {\em does} show reasonable broad,  blue-shifted absorption. By contrast, the spectra on the right have $BI = 0$, despite the fact that they show signs of absorption (and are therefore included in our catalogue). These objects were recognized by the LVQ network and, due to the disagreement between the BI and LVQ verdicts, visually inspected for final classification.

One last note of caution concerns the rates of false positives and negatives among objects that were {\em not} visually inspected. While the sample of 2227
BALQSOs that {\em were} inspected visually may be considered to be fairly reliable, the samples of non-inspected objects are not as clean. In particular, since LVQ and BI alone produce false positives at rates of $11.4\%$ and $13.4\%$, respectively, we may expect $1.5\%$ ($0.114\times0.134$) of the 2130 BALQSOs on which they both agree to be false positives. This amounts to roughly 33 expected false positives in our BALQSO catalogue. Conversely, both methods miss approximately $20\%$ of BALQSOs, so they will erroneously agree on a non-BAL classification for $4\%$ ($0.2\times0.2$) of true BALQSOs. This amounts to roughly 85 false negatives, i.e. 85 BALQSOs that are missing from our catalogue.

Because of the many problems encountered when trying to compile BALQSO catalogues, we have decided to produce for the scientific community a meta-catalogue which includes our whole DR5 parent sample used in this work instead of just a BALQSO catalogue\footnote{An electronic version can be found at \url{http://www.astro.soton.ac.uk/~simo} and on the VizieR server}. The first ten entries of this meta-catalogue are presented in Table \ref{tab:1}. For each QSO in our parent sample we provide all tags that we have found to be useful in creating our own hybrid-LVQ BALQSO catalogue.
\begin{table*}
\begin{centering}
\caption{First 10 objects from our DR5 meta-catalogue. The column names are the same as those used by the SDSS team with the exception of the last 2. \texttt{LVQ\_tag} is set to 1 if the neural network regarded the QSO as a BALQSO, 0 of not. \texttt{Final\_tag} is set to 1 if the QSO is considered as a BALQSO by our hybrid-LVQ method. The BIs have been taken from \protect\cite{gibson08}. The \texttt{ts\_t\_qso} and \texttt{ts\_t\_hiz} columns represent Low-z Quasar selection flag and High-z Quasar selection flag respectively as defined by the SDSS team (\protect\citealt{dr5}). The catalogue can be found in electronic format from \url{http://www.astro.soton.ac.uk/~simo} and the VizieR server.}
\begin{tabular}{c c c c c c c c c c}
\hline
 SDSS Name & RA   & DEC  & z & \texttt{M\_i} & \texttt{ts\_t\_qso} & \texttt{ts\_t\_hiz} & BI & \texttt{LVQ\_tag} & \texttt{Final\_tag}  \\
           & deg. & deg. &   & $M_{i}$      &                   &                   &    & $km~s^{-1}$      &                     \\
\hline
000006.53$+$003055.2 & 0.027228 & 0.515349   & 1.8227 & -25.100 & 0 & 0 & 0    & 0 & 0 \\
000008.13$+$001634.6 & 0.033898 & 0.276304   & 1.8365 & -25.738 & 0 & 0 & 0    & 0 & 0 \\
000009.38$+$135618.4 & 0.039088 & 13.938447  & 2.24   & -27.419 & 1 & 0 & 0    & 0 & 0 \\
000009.42$-$102751.9 & 0.039269 & -10.464428 & 1.8442 & -26.459 & 1 & 0 & 0    & 0 & 0 \\
000013.80$-$005446.8 & 0.057505 & -0.913004  & 1.8361 & -25.648 & 0 & 0 & 0    & 1 & 1 \\
000014.82$-$011030.6 & 0.061778 & -1.175193  & 1.8902 & -26.149 & 0 & 0 & 0    & 0 & 0 \\
000015.47$+$005246.8 & 0.064492 & 0.87968    & 1.8476 & -26.017 & 0 & 0 & 0    & 0 & 0 \\
000030.37$-$002732.4 & 0.126576 & -0.459005  & 1.803  & -25.368 & 0 & 0 & 0    & 0 & 0 \\
000038.65$+$011426.3 & 0.161078 & 1.24064    & 1.8352 & -25.171 & 0 & 0 & 2144 & 1 & 1 \\
000038.99$-$001803.9 & 0.162498 & -0.301102  & 2.1224 & -26.673 & 1 & 0 & 0    & 0 & 0 \\
\hline 
\end{tabular}
\\ 
\label{tab:1}
\end{centering}
\end{table*}

\section{Discussions}

\subsection{The classification of borderline cases}

\begin{figure*}
\centering
\includegraphics[height=\textwidth, angle=90]{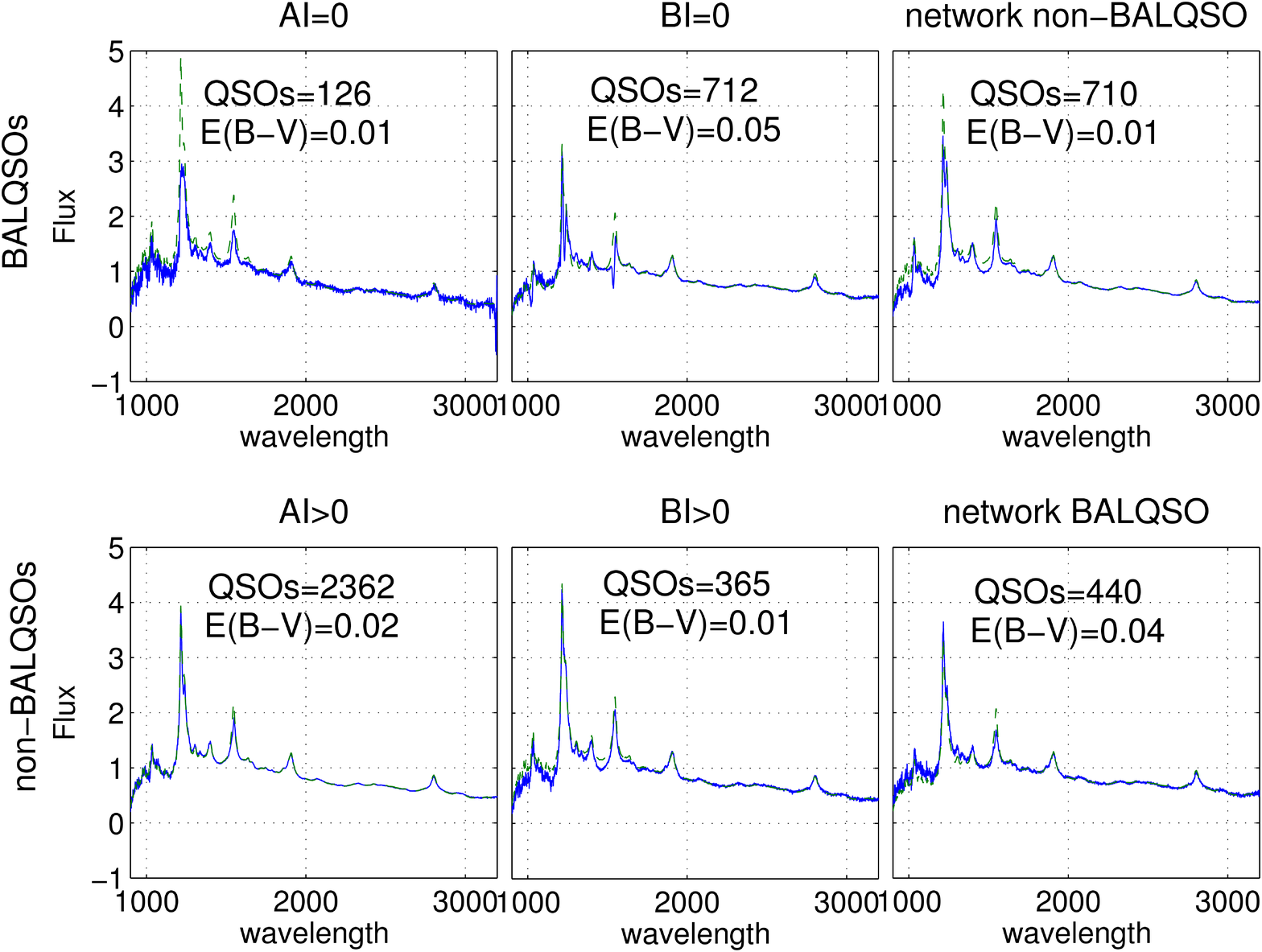}	
\caption{Various composites created in order to examine our BALQSO classification parameter space. The solid blue line displays composites created by selecting different QSOs in our classification parameter space. The dashed green line is a composite created with $AI = 0$ ${\rm km~s^{-1}}$ QSOs after being normalised at 1750\AA\ and de-reddened to match the composites in each panel.}
\label{fig:4}
\end{figure*}

In this section we highlight the difficulties in compiling BALQSO samples using composites derived from our QSO meta-catalogue as shown in Fig. \ref{fig:4}. Each panel displays the non-BALQSO composite (shown in dashed green) normalised to 1750~\AA\ and reddened to match the other composites shown in each
panel (solid blue lines), which were created by selecting relevant QSO sub-sets culled from the meta-catalogue. In the top, row we show composites from QSOs in our parent sample which were finally classified as BALQSOs by our hybrid-LVQ method, subdivided into objects with $AI=0$ ${\rm km~s^{-1}}$ (top-left), $BI=0$ ${\rm km~s^{-1}}$ (top middle) and LVQ non-BAL. All of these composites show clear signatures of absorption blueward of C~{\sc iv}. We note that, for consistency, we have only used objects already included in SDSS DR3 in constructing the composites shown in the left panels, since only these have AI values calculated by \cite{trump06}. 

The composite in the upper left panel comprises objects with $AI=0$ ${\rm km~s^{-1}}$ (and therefore also $BI=0$ ${\rm km~s^{-1}}$) that were classified as BALQSO by the LVQ network and subsequently confirmed as BALQSOs visually. Although only 126 objects were used for the creation of this composite, the absorption near C~{\sc iv} and the slightly truncated emission line are clear BALQSO signatures. These are mostly BALQSOs whose troughs are smaller that $1000$ ${\rm km~s^{-1}}$ (and hence with  $AI=0$ ${\rm km~s^{-1}}$ ). 

The BALQSO composite in the top middle panel was created from objects with $BI=0$ ${\rm km~s^{-1}}$, but subsequently identified as BALQSOs by the LVQ neural network. This composite shows the largest amount of reddening and a fairly narrow and deep absorption blueward of C~{\sc iv}. Such narrow features will, by definition, be classified as non-BALs using the BI metric since they lie within $3000$ ${\rm km~s^{-1}}$ of the line centre.. In the upper right panel we show the composite formed from QSOs with $BI>0$ ${\rm km~s^{-1}},$ identified as non-BALs by the LVQ network, but finally included in the catalogue on the basis of visual inspection. This composite shows a strong, broad, absorption line blueward of C~{\sc iv}, but very little reddening. Furthermore, since the composites in both the top right and top middle panels contain similar numbers of QSOs, it is evident that neither method alone is as reliable in identifying BALQSOs as one might hope. All of the information needed to recreate these composites can be found in our meta-catalogue by querying on various tags. 

The bottom row of Fig. \ref{fig:4} compares those composites comprising QSOs classified as BALQSO candidates using only a single metric that were then re-classified as non-BALs after visual inspection. In detail, the composite at bottom left is created from objects with $AI>0$ ${\rm km~s^{-1}}$, but finally classified as non-BALs by our hybrid-LVQ method. There is virtually no evidence for absorption in this composite spectrum, despite all of the 2362 comprising this composite having $AI>0$ ${\rm km~s^{-1}}$. This again points out the problematic nature of the AI metric for BALQSO selection purposes.

The middle bottom panel displays a composite created from objects having $BI>0$ ${\rm km~s^{-1}}$ (and therefore also $AI>0$ ${\rm km~s^{-1}}$) and a non-BALQSO LVQ tag, finally classified by us as a non-BALQSO. This composite looks very similar to the one on the top right, which was already discussed above. Here again, we see a fairly strong, smooth and broad ($> 3000 {\rm km~s^{-10}}$) absorption feature associated with C~{\sc iv}. The similarity between these two composites is not entirely unexpected, since all of the objects forming them had the same automated classifications (positive BI and a non-LVQ tag), and thus differed only in the outcome of the visual inspection step. Since disagreement between BI and LVQ is most likely to happen for difficult borderline cases, we should certainly expect some mis-classifications and thus overlap between the two sub-sets of QSOs represented by these composites. However, closer inspection does reveal some significant differences between the composites that point to the subtle, but consistent absorption line properties that were obviously picked by the visual classification step. For example, the peaks of the C~{\sc iv} and Lyman-$\alpha$ lines are lower in the top right composite than in the bottom middle one, and only the top right one shows clear evidence of absorption eating into the blue wing of the C~{\sc iv} line (compared to the non-BAL composite). Moreover, even though both composites show some evidence for absorption affecting the bluest part of the spectrum -- shortward of Lyman alpha, and particularly around the Lyman beta and O~{\sc vi} blend near 1030~\AA\ -- this absorption is stronger in the top right panel. Finally, the broad absorption trough associated with C~{\sc iv} in the bottom middle panel is suspiciously symmetric between $2000~{\rm km~s^{-1}}$ and $20,000~{\rm km~s^{-1}}$, the limits within which the BI is calculated. This may indicate that this trough is formed from the superposition of many narrow lines that may or may not be associated with the traditional BAL-flow.

All of these differences are consistent with the idea that the objects represented in the top right panel (which is included in our final BALQSO catalogue) are more likely to be genuine BALQSOs than those represented in the bottom middle panel (which are not included in our final catalogue). However, there is no escaping the fact that the differences are extremely subtle and that a definitive classification scheme for such borderline cases remains elusive. This conclusion is supported by a visual re-inspection of all of the objects contained in these two sub-samples: while we generally remain happy with our classifications as "best-bet estimates", it is clear that in many cases a definitive classification is impossible. Since the 365 borderline cases represent 11\% of the \cite{gibson08} sample, we caution that there is a systematic uncertainty of $\sim$11\% on the BALQSO fraction suggested by even the best presently available classification schemes. This is one of the key reasons we have decided to provide the community with all of the meta-data we have used in constructing our own catalogue.

The last figure on the bottom right displays a composite created by selecting QSOs originally classified as BALQSOs by our neural network but re-classified as non-BALQSOs during the visual inspection phase (these objects all had  $BI=0$ ${\rm km~s^{-1}}$ by definition or they would have not been inspected visually). The composite here is somewhat redder than the non-BAL (E(B - V ) = 0.04), but no clear signatures of absorption are present. This highlights the importance of a visual inspection phase when constructing BALQSO catalogues.

To summarise, it is clear that no single metric (or visual intervention) is adequate in deriving both complete and clean samples of BALQSOs at the moment, so a variety of complementary metrics should instead be employed. Our own experience with unsupervised and supervised learning networks shows that, even though much of the classification work may indeed be automated, human intervention is not only useful, it is often a necessity when dealing with classification involving “not so clearly defined” training samples.

\subsection{The effect of S/N}

\begin{figure}
\centering
\includegraphics[width=0.5\textwidth, height=8cm]{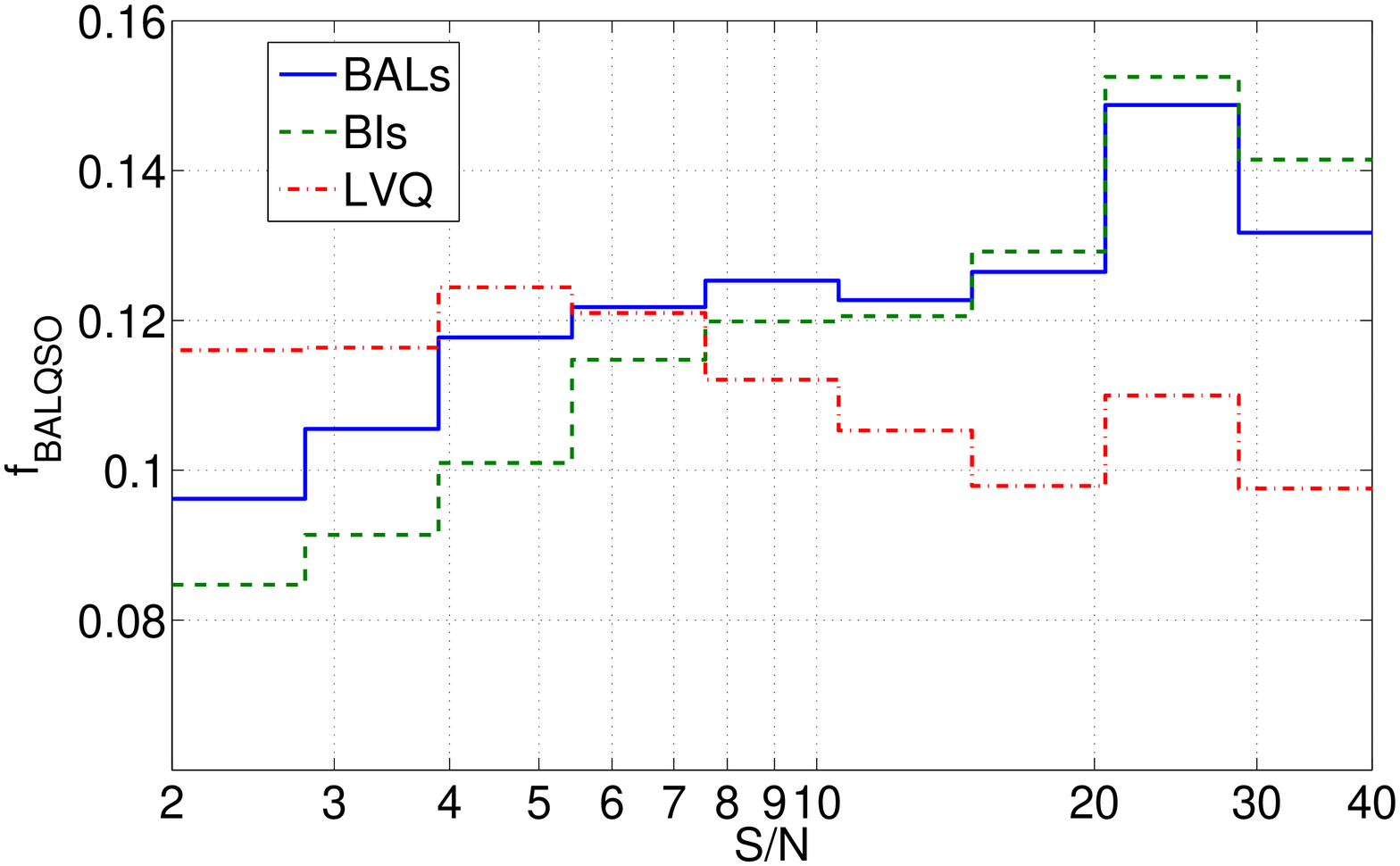}	
\caption{$f_{BALQSO}$ as a function of signal-to-noise calculated in the same way as \protect\cite{gibson08} for BI-selected QSOs, hybrid-LVQ selected QSOs and the final BALQSOs included in our final hybrid-LVQ catalogue.}
\label{fig:6}
\end{figure}

Fig. \ref{fig:6} shows $f_{BALQSO}$ as a function of signal-to-noise for BI-selected QSOs, LVQ selected QSOs and our final BALQSO fraction using our hybrid-LVQ method. The same trend as that found by \cite{gibson08} is evident for BI selected BALQSOs: $f_{BALQSO}$ steadily increases from $\approx 9\%$ in low signal-to-noise data up to $15\%$ in high signal-to-noise data. We suspect that this is because in low signal-to-noise data even relatively small random fluctuations in a shallow BAL trough can trigger the zero reset in the BI calculation and can thus result in $BI=0$ ${\rm km~s^{-1}}$. We note that this would not necessarily be the case if BALs were identified using a more sophisticated metric than the BI to isolate the BAL. We cannot rule out, however, that the apparent trend in the BAL fraction with S/N has a more interesting cause, such as an underlying trend with redshift or luminosity (i.e. the BAL fraction may be higher among low-redshift and/or high-luminosity QSOs, which would also have higher S/N spectra, on average). However, the simpler and more mundane explanation -- that the trend is primarily due to the difficulty in identifying BAL features in low-S/N spectra -- seems far more likely. We have also visually inspected some of the objects with high BI that are not included in our final catalogue and conclude that these are cases where the BI must have been calculated incorrectly and should have been set to $BI=0$ ${\rm km~s^{-1}}$.

By contrast, the BALQSO fraction produced by LVQ alone at high S/N levels is roughly constant, and slightly lower than the fraction suggested by the BI or indicated by our final catalogue. Thus the maximum efficiency of LVQ (when working on high-quality spectra) is comparable to, but slightly less, than that of the BI. Fig. \ref{fig:6} also shows that the LVQ-suggested BALQSO fraction actually {\em increases} towards the lowest S/N levels. Given that the number of low-S/N BALQSOs suggested by LVQ alone is actually higher than that in our final catalogue, and that every LVQ-selected BALQSO candidate was either included in the catalogue or rejected as a false positive via visual inspection, this implies that LVQ has a tendency to classify low-S/N spectra as BALQSOs, leading to a higher false positive rate in this limit. This is not entirely unexpected and actually means that LVQ and BI selections are highly complementary methods when
applied across the full range of S/N levels.

\section{Conclusions}

We have used a recently developed technique for identifying broad absorption lines in quasar spectra to compile a more robust and complete BALQSOs catalogue. Our technique is based on a combination of the traditional ``balnicity index'', a simple neural network and visual inspection of borderline cases and is designed to produce BALQSO samples that are more complete than purely BI-based ones, while still avoiding a high incidence of false positives. Our final catalogue covers the redshift range $1.7 < z < 4.2$ and contains 3552 BALQSOs, corresponding to a raw fraction of $\approx 12.5\%$ of the SDSS DR5 QSOs parent sample with a false positive rate of $\sim$11\%. In the process of constructing a robust BALQSO catalogue, we have explored in detail the classification parameter space for BALQSOs and highlighted the difficulties in BALQSO classification using single metrics. We have also constructed -- and make available -- a meta-catalogue that contains all of the information needed to recreate our BALQSO catalogue from its much larger QSO parent sample, or to create alternative BALQSO samples using different selection criteria. In addition, all of the composite spectra shown in this paper will be made publicly available. Meta-catalogues provide an elegant solution to problems encountered regarding subjectivity and transparency (\citealt{hogg}), in particular when dealing with ``ill-defined'' astronomical objects such as BALQSOs.

\section*{Acknowledgements}
The authors wish to thank the anonymous referee for constructive and useful feedback. This work is supported at the University of Southampton and University of Leicester by the Science and Technology Facilities Council (STFC). Funding for the SDSS and SDSS-II has been provided by the Alfred P. Sloan Foundation, the Participating Institutions, the National Science Foundation, the U.S. Department of Energy, the National Aeronautics and Space Administration, the Japanese Monbukagakusho, the Max Planck Society, and the Higher Education Funding Council for England. The SDSS Web Site is http://www.sdss.org/. The SDSS is managed by the Astrophysical Research Consortium for the Participating Institutions. The Participating Institutions are the American Museum of Natural History, Astrophysical Institute Potsdam, University of Basel, University of Cambridge, Case Western Reserve University, University of Chicago, Drexel University, Fermilab, the Institute for Advanced Study, the Japan Participation Group, Johns Hopkins University, the Joint Institute for Nuclear Astrophysics, the Kavli Institute for Particle Astrophysics and Cosmology, the Korean Scientist Group, the Chinese Academy of Sciences (LAMOST), Los Alamos National Laboratory, the Max-Planck-Institute for Astronomy (MPIA), the Max-Planck-Institute for Astrophysics (MPA), New Mexico State University, Ohio State University, University of Pittsburgh, University of Portsmouth, Princeton University, the United States Naval Observatory, and the University of Washington.

\bibliographystyle{mn2e}
\bibliography{dr5_paper}

\appendix
\label{lastpage}

\end{document}